\documentclass[preprint,prX]{revtex4}
\usepackage{bm}
\usepackage{amsfonts}
\usepackage{amsmath}
\usepackage{amssymb}
\usepackage{graphicx}

\begin{document}

\title[Neutrino emission from triplet paring]{Neutrino emission from triplet
pairing of neutrons in neutron stars}
\author{L. B. Leinson}
\affiliation{Institute of Terrestrial Magnetism, Ionosphere and Radio Wave Propagation
RAS, 142190 Troitsk, Moscow Region, Russia}

\begin{abstract}
Neutrino emission due to the pair breaking and formation processes in the
bulk triplet superfluid in neutron stars is investigated with taking into
account of anomalous weak interactions. We consider the problem in the BCS
approximation discarding Fermi-liquid effects. In this approach we derive
self-consistent equations for anomalous vector and axial-vector vertices of
weak interactions taking into account the $^{3}P_{2}-~^{3}F_{2}$ mixing.
Further we simplify the problem and consider the pure $^{3}P_{2}$ pairing
with $m_{j}=0$, as is adopted in the minimal cooling paradigm. As was
expected because of current conservation we have obtained a large
suppression of the neutrino emissivity in the vector channel. More exactly,
the neutrino emission through the vector channel vanishes in the
nonrelativistic limit $V_F=0$. The axial channel is also found to be
moderately suppressed. The total neutrino emissivity is suppressed by a
factor of $1.9\times10^{-1}$ relative to original estimates using bare weak
vertices.
\end{abstract}

\startpage{1}
\maketitle

\section{Introduction}

Thermal excitations in superfluid baryon matter of neutron stars, in the
form of broken Cooper pairs, can recombine into the condensate by emitting
neutrino pairs via neutral weak currents \cite{FRS76}. It is generally
accepted that, for temperatures near the associated superfluid critical
temperatures, emission from pair breaking and formation (PBF) processes
dominates the neutrino emissivities in many cases. Recently \cite{LP06}, it
has been found however that the existing theory of PBF processes based on
the bare weak vertices violates conservation of vector weak current. Correct
evaluations including anomalous interactions has shown the neutrino emission
by a nonrelativistic singlet superfluid is substantially suppressed.
Consistent estimates of the inhibition factor can be found in Refs. 
\cite{LP06}-\cite{L09}. The suppression of neutrino emissivity from the 
$^{1}S_{0}$ PBF processes was studied also in Refs. \cite{Reddy}-\cite{SR}, 
although these are controversial (see
discussion in Refs. \cite{L08}, \cite{L09}).

Quenching of the neutrino emission found in the case of $^{1}S_{0}$ pairing
leads to higher temperatures that can be reached in the crust of an
accreting neutron star. This allows to explain the observed data of
superbursts triggering \cite{Cumming}, \cite{Gupta} which was in dramatic
discrepancy with the previous theory of the crust cooling. Numerical
simulations of the neutron star cooling in the minimal scenario \cite{Page09}
have shown that the suppression of the PBF processes in the crust of a
neutron star has a significant effect at early times ($t<1000$ years) and
results in warmer crusts and increased crust relaxation times.

We now turn to the PBF neutrino emission from the bulk superfluid neutron
matter which is mostly caused by the triplet neutron pairing. Neutrino
energy losses due to the triplet PBF processes have been initially derived
in Ref. \cite{YKL}, ignoring the anomalous weak interactions. From analogy
with the singlet case it is clear that conservation of the vector weak
current is violated in this approach and thus the neutrino emission in the
vector channel, as obtained in Ref. \cite{YKL}, is a subject of
inconsistency \cite{LP06a}. Moreover, in the triplet superfluid, the order
parameter is sensitive also to the axial weak field. Therefore the
self-consistent axial response of the triplet superfluid must incorporate
the anomalous contributions in the same degree of approximation as the
vector response. This effect is not investigated up to now.

In present paper, we perform the corresponding self-consistent calculation.
Formally, our approach is a development of Larkin-Migdal-Leggett theory \cite%
{Larkin}, \cite{Leggett} to the triplet case. However, we discard residual
particle-hole interactions because the Landau parameters are unknown for a
dense asymmetric baryon matter. Another reason is that the influence of the
particle-hole interactions is not very significant in the PBF processes \cite%
{L09}.

The paper is organized as follows. The next section contains some
preliminary notes. We discuss the order parameter and the quasiparticle
propagators for the triplet pair-correlated system with strong interactions.
We also recast the standard gap equation to the form convenient for
consideration of the processes occuring in a vicinity of the Fermi surface.
In Sec. III, we formulate the set of BCS equations for calculation of the
anomalous vertices and correlation functions of the triplet superfluid Fermi
liquid at finite temperature involving a mixing of the $^{3}P_{2}$ and $%
^{3}F_{2}$ channels \cite{Tamagaki}, \cite{Takatsuka}. In Sec. IV, we
present the general expression for the emissivity of the neutron PBF
processes formulated in terms of the imaginary part of the current-current
correlator. The widely used expression for the neutrino emissivity caused by
the triplet pairing of neutrons was obtained in Ref. \cite{YKL} with the aid
of the Fermi golden rule. Therefore before proceeding to the self-consistent
calculation of the neutrino energy losses, in Sec. V, we reproduce this
formula using the calculation technique developed in our paper so that an
apposite comparison with Ref. \cite{YKL} can be made. In Sec. VI, we
consider the anomalous vertices and the self-consistent superfluid response
both in the vector and axial channels. Here we focus on the $^{3}P_{2}$
pairing with $m_{j}=0$, as is adopted in the minimal cooling paradigm \cite%
{Page09}. Finally, in Sec. VII, we evaluate the self-consistent neutrino
energy losses from the PBF processes in the triplet neutron superfluid.
Section VIII contains a short summary of our findings and the conclusion.

In this work we use the standard model of weak interactions, the system of
units $\hbar=c=1,$ and the Boltzmann constant $k_{B}=1$.

\section{Preliminary notes and notation}

\subsection{The order parameter and Green functions.}

The order parameter, $\hat{D}\equiv D_{\alpha\beta}$, arising due to triplet
pairing of quasiparticles, represents a $2\times2$ symmetric matrix in spin
space, $\left( \alpha,\beta=\uparrow,\downarrow\right) $. The spin-orbit
interaction among quasiparticles is known to dominate in the nucleon matter
of a high density. Therefore it is conventional to represent the triplet
order parameter of the system $\hat{D}=\sum_{lm_{j}}\Delta_{lm_{j}}\Phi_{%
\alpha \beta}^{\left( jlm_{j}\right) }$ as a superposition of standard
spin-angle functions of the total angular momentum $\left( j,m_{j}\right) $, 
\begin{equation}
\Phi_{\alpha\beta}^{\left( jlm_{j}\right) }\left( \mathbf{n}\right)
\equiv\sum_{m_{s}+m_{l}=m_{j}}\left( \frac{1}{2}\frac{1}{2}\alpha\beta
|sm_{s}\right) \left( slm_{s}m_{l}|jm_{j}\right) Y_{l,m_{l}}\left( \mathbf{n}%
\right) .  \label{sa}
\end{equation}

For our calculations it will be more convenient to use vector notation which
involves a set of mutually orthogonal complex\ vectors $\mathbf{b}%
_{lm_{j}}\left( \mathbf{n}\right) $ defined as%
\begin{equation}
\mathbf{b}_{lm_{j}}\left( \mathbf{n}\right) =-\frac{1}{2}\mathrm{Tr}\left( 
\hat{g}\bm{\hat{\sigma}}\hat{\Phi}_{jlm_{j}}\right) ~,  \label{blm}
\end{equation}%
where $\bm{\hat{\sigma}}=\left( \hat{\sigma}_{1},\hat{\sigma}_{2},\hat{\sigma%
}_{3}\right) $ are Pauli spin matrices, and $\hat{g}=i\hat{\sigma}_{2}$. We
will use the normalization condition%
\begin{equation}
\int \frac{d\mathbf{n}}{4\pi }\mathbf{b}_{l^{\prime }m_{j}^{\prime }}^{\ast }%
\mathbf{b}_{lm_{j}}=\delta _{ll^{\prime }}\delta _{m_{j}m_{j}^{\prime }}.
\label{lmnorm}
\end{equation}

If the most attractive channel of interactions is assumed in the states with 
$s=1,j=2,l=j\pm1$ (in the case of tensor forces) the order parameter can be
written in the form 
\begin{equation}
\hat{D}\left( \mathbf{n}\right) =\sum_{lm_{j}}\Delta_{lm_{j}}\left( %
\bm{\hat{\sigma}}\mathbf{b}_{lm_{j}}\right) \hat{g}\ .  \label{Dnlm}
\end{equation}
We are mostly interested in the values of quasiparticle momenta \textbf{p}
near the Fermi surface $p\simeq p_{F}$, where the partial gap amplitudes, $%
\Delta_{lm_{j}}\left( p\right) \simeq\Delta_{lm_{j}}\left( p_{F}\right) ,$
are almost constants, and the angular dependence of the order parameter is
represented by the unit vector $\mathbf{n=p}/p$ which defines the polar
angles $\left( \theta,\varphi\right) $ on the Fermi surface.

The ground state (\ref{Dnlm})\ occurring in neutron matter has a relatively
simple structure (unitary triplet) \cite{Tamagaki}, \cite{Takatsuka}: 
\begin{equation}
\sum_{lm_{j}}\Delta_{lm_{j}}\mathbf{b}_{lm_{j}}\left( \mathbf{n}\right)
=\Delta~\mathbf{\bar{b}}\left( \mathbf{n}\right) ~,  \label{bbar}
\end{equation}
where $\Delta$ is a complex constant (on the Fermi surface), and $\mathbf{%
\bar{b}}\left( \mathbf{n}\right) $ is a real vector which we normalize by
the condition 
\begin{equation}
\int\frac{d\mathbf{n}}{4\pi}\bar{b}^{2}\left( \mathbf{n}\right) =1~.
\label{Norm}
\end{equation}
Thus the triplet order parameter can be written as 
\begin{equation}
\hat{D}\left( \mathbf{n}\right) =\Delta\mathbf{\bar{b}}\bm{\hat{\sigma}}%
\hat {g}~.  \label{Dn}
\end{equation}

We will use the adopted graphical notation for the ordinary and anomalous
propagators, as shown in Fig. 1.

\begin{figure}[h]
\includegraphics{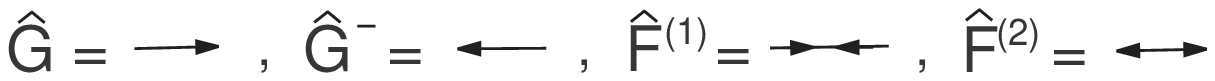}
\caption{Diagrams depicting the ordinary and anomalous propagators of a
quasiparticle.}
\label{fig1}
\end{figure}

The analytic form of the propagators can be found in the standard way \cite%
{AGD}, \cite{Migdal} , using the general form (\ref{Dn}) of the gap matrix.
Since the matter is assumed in thermal equilibrium at some temperature, we
employ the Matsubara calculation technique. Then%
\begin{align}
\hat{G}\left( p_{m},\mathbf{p}\right) & =aG\left( p_{m},\mathbf{p}\right)
\delta _{\alpha \beta }~,\ \ \ \ \ \ \ \hat{G}^{-}\left( p_{m},\mathbf{p}%
\right) =aG^{-}\left( p_{m},\mathbf{p}\right) \delta _{\alpha \beta }~, 
\notag \\
\hat{F}^{\left( 1\right) }\left( p_{m},\mathbf{p}\right) & =aF\left( p_{m},%
\mathbf{p}\right) \mathbf{\bar{b}}\bm{\hat{\sigma}}\hat{g}~,\ \ \ \hat{F}%
^{\left( 2\right) }\left( p_{m},\mathbf{p}\right) =aF\left( p_{m},\mathbf{p}%
\right) \hat{g}\bm{\hat{\sigma}}\mathbf{\bar{b}}~,  \label{GF}
\end{align}%
where $a\simeq 1$ is the usual Green's-function renormalization constant; $%
p_{m}\equiv i\pi \left( 2m+1\right) T$ with $m=0,\pm 1,\pm 2...$ is the
Matsubara's fermion frequency, and the scalar Green's functions are of the
form%
\begin{align}
G\left( p_{m},\mathbf{p}\right) & =\frac{-ip_{m}-\varepsilon _{\mathbf{p}}}{%
p_{m}^{2}+E_{\mathbf{p}}^{2}}~,\ G^{-}\left( p_{m},\mathbf{p}\right) =\frac{%
ip_{m}-\varepsilon _{\mathbf{p}}}{p_{m}^{2}+E_{\mathbf{p}}^{2}}~,  \notag \\
F\left( p_{m},\mathbf{p}\right) & =\frac{-\Delta }{p_{m}^{2}+E_{\mathbf{p}%
}^{2}}~.  \label{GFc}
\end{align}%
Here%
\begin{equation}
\varepsilon _{\mathbf{p}}=\frac{p^{2}}{2M^{\ast }}-\frac{p_{F}^{2}}{2M^{\ast
}}\simeq \frac{p_{F}}{M^{\ast }}(p-p_{F}),  \label{ksi}
\end{equation}%
with $M^{\ast }=p_{F}/V_{F}$ being the effective mass of a quasiparticle. $%
\allowbreak \allowbreak $The quasiparticle energy is given by%
\begin{equation}
E_{\mathbf{p}}\equiv \sqrt{\varepsilon _{\mathbf{p}}^{2}+\frac{1}{2}\mathrm{%
Tr}\hat{D}\left( \mathbf{n}\right) \hat{D}^{\dagger }\left( \mathbf{n}%
\right) }=\sqrt{\varepsilon _{\mathbf{p}}^{2}+\Delta ^{2}\bar{b}^{2}}~,
\label{Ep}
\end{equation}%
where the (temperature-dependent) energy gap, $\Delta \bar{b}\left( \mathbf{n%
}\right) $, is anisotropic. Here the fact is used that, in the absence of
external fields, the gap amplitude $\Delta $ is real.

Green functions of a quasiparticle (\ref{GF}) involve the renormalization
factor $a\simeq1$ independent of $\omega,\mathbf{q},T$ (see e.g. \cite%
{Migdal}). The final outcomes are independent of this factor therefore we
will drop the renormalization factor in order to shorten the equations by
assuming that all the necessary physical values are properly renormalized.

The following notation will be used below. We designate as $L_{X,X}\left(
\omega ,\mathbf{q;p}\right) $ the analytical continuation onto the
upper-half plane of complex variable $\omega $ of the following Matsubara
sums:%
\begin{equation}
L_{XX^{\prime }}\left( \omega _{n},\mathbf{p+}\frac{\mathbf{q}}{2}\mathbf{;p-%
}\frac{\mathbf{q}}{2}\right) =T\sum_{m}X\left( p_{m}+\omega _{n},\mathbf{p+}%
\frac{\mathbf{q}}{2}\right) X^{\prime }\left( p_{m},\mathbf{p-}\frac{\mathbf{%
q}}{2}\right) ~,  \label{LXX}
\end{equation}%
where $X,X^{\prime }\in G,F,G^{-}$, and $\omega _{n}=2i\pi Tn$ with $n=0,\pm
1,\pm 2...$.

It is convenient to divide the integration over the momentum space into
integration over the solid angle and over the energy according to%
\begin{equation}
\int\frac{d^{3}p}{\left( 2\pi\right) ^{3}}...=\rho\int\frac{d\mathbf{n}}{4\pi%
}\frac{1}{2}\int_{-\infty}^{\infty}d\varepsilon_{\mathbf{p}}...  \label{1}
\end{equation}
and operate with integrals over the quasiparticle energy:%
\begin{equation}
\mathcal{I}_{XX^{\prime}}\left( \omega,\mathbf{n,q};T\right) \equiv\frac {1}{%
2}\int_{-\infty}^{\infty}d\varepsilon_{\mathbf{p}}L_{XX^{\prime}}\left(
\omega,\mathbf{p+}\frac{\mathbf{q}}{2}\mathbf{,p-}\frac{\mathbf{q}}{2}%
\right) ~.  \label{IXX}
\end{equation}
These are functions of $\omega$, $\mathbf{q}$ and the direction of the
quasiparticle momentum $\mathbf{p}=p\mathbf{n}$. Here and below $\rho
=p_{F}M^{\ast}/\pi^{2}$ is the density of states near the Fermi surface.

The loop integrals (\ref{IXX}) possess the following properties which can be
verified by a straightforward calculation:%
\begin{equation}
\mathcal{I}_{G^{-}G}=\mathcal{I}_{GG^{-}}~,~\mathcal{I}_{GF}=-\mathcal{I}%
_{FG}~,~\mathcal{I}_{G^{-}F}=-\mathcal{I}_{FG^{-}}~,  \label{Leg}
\end{equation}%
\begin{equation}
\mathcal{I}_{G^{-}F}+\mathcal{I}_{FG}=\frac{\omega}{\Delta}\mathcal{I}_{FF}~,
\label{gf}
\end{equation}%
\begin{equation}
\mathcal{I}_{G^{-}F}-\mathcal{I}_{FG}=-\frac{\mathbf{qv}}{\Delta}\mathcal{I}%
_{FF}~.  \label{ff1}
\end{equation}

For arbitrary $\omega ,\mathbf{q},T$ one can obtain also 
\begin{equation}
\mathcal{I}_{GG^{-}}+\bar{b}^{2}\mathcal{I}_{FF}=A+\frac{\omega ^{2}-\left( 
\mathbf{qv}\right) ^{2}}{2\Delta ^{2}}\mathcal{I}_{FF}~,  \label{FF}
\end{equation}%
where $\mathbf{v}$ is a vector with the magnitude of the Fermi velocity $%
V_{F}$ and the direction of $\mathbf{n}$, and%
\begin{equation}
A\left( \mathbf{n}\right) \equiv \left[ \mathcal{I}_{G^{-}G}\left( \mathbf{n}%
\right) +\bar{b}^{2}\left( \mathbf{n}\right) \mathcal{I}_{FF}\left( \mathbf{n%
}\right) \right] _{\omega =0,\mathbf{q}=0}~.  \label{A}
\end{equation}

In the case of triplet superfluid the key role in the response theory
belongs to the loop integrals $\mathcal{I}_{FF}$ and $\left( \mathcal{I}%
_{GG}\pm \bar{b}^{2}\mathcal{I}_{FF}\right) $. For further usage we indicate
the properties of thise functions in the case of $\omega >0$ and $\mathbf{q}%
\rightarrow 0$. A straightforward calculation yields%
\begin{equation}
\mathcal{I}_{FF}\left( \omega ,q=0\right) =-2\Delta ^{2}\int_{0}^{\infty }%
\frac{d\varepsilon }{E}\frac{1}{\left( \omega +i0\right) ^{2}-4E^{2}}\tanh 
\frac{E}{2T}~,  \label{FFq0}
\end{equation}%
and%
\begin{equation}
\left( \mathcal{I}_{GG}+\bar{b}^{2}\mathcal{I}_{FF}\right) _{q\rightarrow
0}=0~,  \label{GGpFF}
\end{equation}%
\begin{equation}
\left( \mathcal{I}_{GG}-\bar{b}^{2}\mathcal{I}_{FF}\right) _{q\rightarrow
0}=-2\bar{b}^{2}\mathcal{I}_{FF}\left( \omega ,\mathbf{0}\right) ~.
\label{GGmFF}
\end{equation}%
The imaginary part of $\mathcal{I}_{FF}$ arises from the poles of the
integrand in Eq. (\ref{FFq0}) at $\omega =\pm 2E$:%
\begin{equation}
\operatorname{Im}\mathcal{I}_{FF}\left( \omega >0,q=0\right) =\Theta \left( \omega
^{2}-4\bar{b}^{2}\Delta ^{2}\right) \frac{\pi \Delta ^{2}}{\omega \sqrt{%
\omega ^{2}-4\bar{b}^{2}\Delta ^{2}}}\tanh \frac{\omega }{4T}~.
\label{IFFq0}
\end{equation}%
where $\Theta \left( x\right) $ is Heaviside step function.

\subsection{Gap equation}

The block of the interaction diagrams irreducible in the channel of two
quasiparticles, $\Gamma _{\alpha \beta ,\gamma \delta }$, is usually
generated by the expansion over spin-angle functions (\ref{sa}). Using the
vector notation, the most attractive channel of pairing interactions with $%
j=2$ can be written as 
\begin{equation}
\rho \Gamma _{\alpha \beta ,\gamma \delta }\left( \mathbf{p,p}^{\prime
}\right) =-\sum_{l^{\prime }lm_{j}}V_{ll^{\prime }}\left( p,p^{\prime
}\right) \left( \mathbf{b}_{lm_{j}}(\mathbf{n})\bm{\hat{\sigma}}\hat{g}%
\right) _{\alpha \beta }\left( \hat{g}\bm{\hat{\sigma}}\mathbf{b}_{l^{\prime
}m_{j}}^{\ast }(\mathbf{n}^{\prime })\right) _{\gamma \delta }~,
\label{ppint}
\end{equation}%
$\allowbreak $ where $V_{ll^{\prime }}\left( p,p^{\prime }\right) $ are the
corresponding interaction amplitudes, and $|l-l^{\prime }|\leq 2$ in the
case of tensor forces.

In vector notation the set of equations for the triplet partial amplitudes $%
\Delta _{lm_{j}}$ is of the form 
\begin{equation}
\Delta _{lm_{j}}\left( p\right) =-\sum_{l^{\prime }}\frac{1}{2\rho }\int
dp^{\prime }p^{\prime 2}V_{ll^{\prime }}\left( p,p^{\prime }\right) \Delta
\left( p^{\prime }\right) \int \frac{d\mathbf{n}^{\prime }}{4\pi }\mathbf{b}%
_{l^{\prime }m_{j}}^{\ast }(\mathbf{n}^{\prime })\mathbf{\bar{b}}(\mathbf{n}%
^{\prime })T\sum_{m}\frac{1}{p_{m}^{2}+E_{\mathbf{p}^{\prime }}^{2}}~.
\label{gapset}
\end{equation}%
where 
\begin{equation}
\mathbf{\bar{b}}\left( \mathbf{n}\right) =\frac{1}{\Delta }%
\sum_{lm_{j}}\Delta _{lm_{j}}\mathbf{b}_{lm_{j}}\left( \mathbf{n}\right) 
\text{,}  \label{bb}
\end{equation}%
as defined in Eq. (\ref{bbar}). These equations can be reduced to the
standard form \cite{Takatsuka} with the aid of the identity%
\begin{equation}
T\sum_{m}\frac{1}{p_{m}^{2}+E_{\mathbf{p}^{\prime }}^{2}}\equiv {\frac{1}{2E(%
\mathbf{p}^{\prime })}}\tanh {\frac{E(\mathbf{p}^{\prime })}{2T}~},
\label{sum}
\end{equation}%
and the relation%
\begin{equation}
\frac{1}{2}\mathrm{Tr}\left( \hat{\Phi}_{jlm_{j}}\hat{\Phi}_{jl^{\prime
}m_{j}}^{\ast }\right) =\allowbreak \mathbf{b}_{lm_{j}}\left( \mathbf{n}%
\right) \cdot \mathbf{b}_{l^{\prime }m_{j}}^{\ast }\left( \mathbf{n}\right)
~.  \label{Tr}
\end{equation}

We are interested in the processes occuring in a vicinity of the Fermi
surface. Therefore we now recast the gap equation to the more convenient
form. We notice that%
\begin{equation}
\frac{1}{p_{m}^{2}+E_{\mathbf{p}}^{2}}\equiv G\left( p_{m},\mathbf{p}\right)
G^{-}\left( p_{m},\mathbf{p}\right) +\bar{b}^{2}F\left( p_{m},\mathbf{p}%
\right) F\left( p_{m},\mathbf{p}\right) \ ,  \label{GGm}
\end{equation}
i.e. Eq. (\ref{gapset}) can be written as 
\begin{align}
\Delta_{lm_{j}}\left( p\right) & =-\frac{1}{2\rho}\sum_{l^{\prime}}\int
dp^{\prime}p^{\prime2}V_{ll^{\prime}}\left( p,p^{\prime}\right) \Delta\left(
p^{\prime}\right) \int\frac{d\mathbf{n}^{\prime}}{4\pi }\mathbf{b}%
_{l^{\prime}m_{j}}^{\ast}(\mathbf{n}^{\prime})\mathbf{\bar{b}}(\mathbf{n}%
^{\prime})  \notag \\
& \times T\sum_{m}\left[ G\left( p_{m},\mathbf{p}^{\prime}\right)
G^{-}\left( p_{m},\mathbf{p}^{\prime}\right) +\bar{b}^{2}F\left( p_{m},%
\mathbf{p}^{\prime}\right) F\left( p_{m},\mathbf{p}^{\prime}\right) \right]
\ .  \label{gap11}
\end{align}

To get rid of the integration over the regions far from the Fermi surface we
renormalize the interaction as suggested in Ref. \cite{Leggett}: we define%
\begin{equation}
V_{ll^{\prime}}^{\left( r\right) }\left( p,p^{\prime};T\right)
=V_{ll^{\prime}}\left( p,p^{\prime}\right) -V_{ll^{\prime}}\left(
p,p^{\prime}\right) \left( GG^{-}\right) _{n}V_{ll^{\prime}}^{\left(
r\right) }\left( p,p^{\prime};T\right) \ ,  \label{Vr}
\end{equation}
where the loop $\left( GG^{-}\right) _{n}$ is evaluated in the normal
(nonsuperfluid) state. In terms of $V_{ll^{\prime}}^{\left( r\right) }$ the
gap equation becomes%
\begin{align}
\Delta_{lm_{j}}\left( p\right) & =-\frac{1}{2\rho}\sum_{l^{\prime}}\int
dp^{\prime}p^{\prime2}V_{ll^{\prime}}^{\left( r\right) }\left( p,p^{\prime
}\right) \Delta\left( p^{\prime}\right) \int\frac{d\mathbf{n}^{\prime}}{4\pi}%
\mathbf{b}_{l^{\prime}m_{j}}^{\ast}(\mathbf{n}^{\prime})\mathbf{\bar {b}}(%
\mathbf{n}^{\prime})  \notag \\
& \times T\sum_{m}\left[ GG^{-}-\left( GG^{-}\right) _{n}+\bar{b}^{2}FF%
\right] _{p_{m},\mathbf{p}^{\prime}}\ .  \label{gap12}
\end{align}
and we may everywhere substitute $V_{ll^{\prime}}^{\left( r\right) }$ for $%
V_{ll^{\prime}}$ provided that at the same time we understand by $GG^{-}$
element the subtracted quantity $GG^{-}-\left( GG^{-}\right) _{n}$ [$\left(
GG^{-}\right) _{n}$ is to be evaluated for $\omega=0,\mathbf{q}=0$ in all
cases]. From now we will do this and drop the superscript $r$ on $%
V_{ll^{\prime}}^{\left( r\right) }$.

Since the function $GG^{-}+\bar{b}^{2}FF$ decreases rapidly along with a
distance from the Fermi surface, we may replace Eq. (\ref{gap12}) with%
\begin{equation}
\Delta_{lm_{j}}=-\frac{1}{\rho}\sum_{l^{\prime}}V_{ll^{\prime}}\Delta\int 
\frac{d\mathbf{n}}{4\pi}\mathbf{b}_{l^{\prime}m_{j}}^{\ast}(\mathbf{n})%
\mathbf{\bar{b}}(\mathbf{n})\frac{1}{2}\int dpp^{2}T\sum_{m}\left[ GG^{-}+%
\bar{b}^{2}FF\right] _{p_{m},\mathbf{p}}\ ,  \label{vgap}
\end{equation}
assuming that in the narrow vicinity of the Fermi surface the smooth
functions $\Delta_{lm_{j}}\left( p\right) ,~V_{ll^{\prime}}\left(
p,p^{\prime }\right) ,~\Delta\left( p^{\prime}\right) $ may be replaced with
constants: $\Delta\left( p\right) \simeq\Delta\left( p_{F}\right)
\equiv\Delta$, ect..

The function (\ref{A}) is now to be understood as%
\begin{equation}
A\left( \mathbf{n}\right) \rightarrow \left[ \mathcal{I}_{G^{-}G}-\mathcal{I}%
_{\left( G^{-}G\right) _{n}}+\bar{b}^{2}\mathcal{I}_{FF}\right] _{\omega =0,%
\mathbf{q}=0}~,  \label{Ar}
\end{equation}%
and the gap equations (\ref{vgap}) become:%
\begin{equation}
\Delta _{lm_{j}}=-\Delta \sum_{l^{\prime }}V_{ll^{\prime }}\int \frac{d%
\mathbf{n}}{4\pi }\mathbf{b}_{l^{\prime }m_{j}}^{\ast }(\mathbf{n})\mathbf{%
\bar{b}}(\mathbf{n})A\left( \mathbf{n}\right) \ .  \label{gapeq}
\end{equation}%
The function (\ref{Ar}) can be found explicitly after performing the
Matsubara's summation:%
\begin{equation}
A\left( \mathbf{n}\right) =\frac{1}{4}\int_{-\infty }^{\infty }d\varepsilon
\left( \frac{1}{\sqrt{\varepsilon ^{2}+\Delta ^{2}\bar{b}^{2}}}\tanh \frac{%
\sqrt{\varepsilon ^{2}+\Delta ^{2}\bar{b}^{2}}}{2T}-\frac{1}{\varepsilon }%
\tanh \frac{\varepsilon }{2T}\right) ~.  \label{An}
\end{equation}

\section{Effective vertices and the correlation functions}

$\allowbreak $The field interaction with a superfluid should be described
with the aid of four effective three-point vertices shown in Fig. 2.

\begin{figure}[h]
\includegraphics{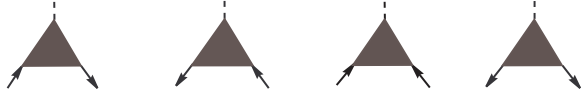}
\caption{Diagrams of the ordinary and anomalous vertices for the
quasiparticle interacting with the external field shown by the dash line.}
\label{fig2}
\end{figure}

There are two ordinary effective vertices corresponding to creation of a
particle and a hole by the field that differ by direction of fermion lines.
We denote these $2\times 2$ matrices as $\hat{\tau}\left( \mathbf{n;}\omega ,%
\mathbf{q}\right) \equiv \tau _{\alpha \beta }\left( \mathbf{n;}\omega ,%
\mathbf{q}\right) $ and $\hat{\tau}^{-}\left( \mathbf{n;}\omega ,\mathbf{q}%
\right) \equiv \tau _{\beta \alpha }\left( -\mathbf{n;}\omega ,\mathbf{q}%
\right) $, respectively. The anomalous vertices correspond to creation of
two particles or two holes. We denote these matrices as $\hat{T}^{\left(
1\right) }\left( \mathbf{n;}\omega ,\mathbf{q}\right) $ and $\hat{T}^{\left(
2\right) }\left( \mathbf{n;}\omega ,\mathbf{q}\right) $, respectively.

Given by the sum of the ladder-type diagrams \cite{Larkin}, the anomalous
vertices are to satisfy the Dyson's equations symbolically depicted by the
graphs in Fig. 3.

\begin{figure}[h]
\includegraphics{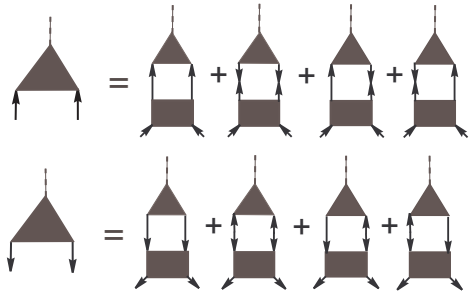}
\caption{Dyson's equations for the anomalous vertices. The shaded rectangle
represents the pairing interaction.}
\label{fig3}
\end{figure}

Analytically the equations reduce to the following (we omit for brevity the
dependence of functions on $\omega $ and $\mathbf{q}$\textbf{)}: 
\begin{align}
T_{\alpha \beta }^{\left( 1\right) }\left( \mathbf{n}\right) &
=\sum_{lm_{j}}\left( \bm{\hat{\sigma}}\mathbf{b}_{lm_{j}}(\mathbf{n})\hat{g}%
\right) _{\alpha \beta }\sum_{l^{\prime }}V_{ll^{\prime }}  \notag \\
& \times \int \frac{d\mathbf{n}^{\prime }}{8\pi }\mathrm{Tr}\left[ \mathcal{I%
}_{GG^{-}}\hat{g}\left( \bm{\hat{\sigma}}\mathbf{b}_{l^{\prime }m_{j}}^{\ast
}\right) \hat{T}^{\left( 1\right) }-\mathcal{I}_{FF}\left( \bm{\hat{\sigma}}%
\mathbf{b}_{l^{\prime }m_{j}}^{\ast }\right) \left( \bm{\hat{\sigma}}\mathbf{%
\bar{b}}\right) \hat{g}\hat{T}^{\left( 2\right) }\left( \bm{\hat{\sigma}}%
\mathbf{\bar{b}}\right) \right.  \notag \\
& \left. -\mathcal{I}_{GF}\left( \bm{\hat{\sigma}}\mathbf{\bar{b}}\right)
\left( \bm{\hat{\sigma}}\mathbf{b}_{l^{\prime }m_{j}}^{\ast }\right) \hat{%
\tau}+\mathcal{I}_{FG^{-}}\left( \bm{\hat{\sigma}}\mathbf{b}_{l^{\prime
}m_{j}}^{\ast }\right) \left( \bm{\hat{\sigma}}\mathbf{\bar{b}}\right)
\left( \hat{g}\hat{\tau}^{-}\hat{g}\right) \right] _{\mathbf{n}^{\prime }}~,
\label{T1eq}
\end{align}%
\begin{align}
T_{\alpha \beta }^{\left( 2\right) }\left( \mathbf{n}\right) &
=\sum_{lm_{j}}\left( \hat{g}\bm{\hat{\sigma}}\mathbf{b}_{lm_{j}}^{\ast }(%
\mathbf{n})\right) _{\alpha \beta }\sum_{l^{\prime }}V_{ll^{\prime }}  \notag
\\
& \times \int \frac{d\mathbf{n}^{\prime }}{8\pi }\mathrm{Tr}\left[ \mathcal{I%
}_{G^{-}G} \left( \bm{\hat{\sigma}}\mathbf{b}_{l^{\prime }m_{j}}\right) \hat{%
g}\hat{T}^{\left( 2\right) }-\mathcal{I}_{FF}\left( \bm{\hat{\sigma}}\mathbf{%
b}_{l^{\prime }m_{j}}\right) \left( \bm{\hat{\sigma}}\mathbf{\bar{b}}\right) 
\hat{T}^{\left( 1\right) }\hat{g}\left( \bm{\hat{\sigma}}\mathbf{\bar{b}}%
\right) \right.  \notag \\
& \left. +\mathcal{I}_{G^{-}F}\left( \bm{\hat{\sigma}}\mathbf{b}_{l^{\prime
}m_{j}}\right) \hat{g}\hat{\tau}^{-}\hat{g}\left( \bm{\hat{\sigma}}\mathbf{%
\bar{b}}\right) -\mathcal{I}_{FG}\left( \bm{\hat{\sigma}}\mathbf{b}%
_{l^{\prime }m_{j}}\right) \left( \bm{\hat{\sigma}}\mathbf{\bar{b}}\right) 
\hat{\tau}\right] _{\mathbf{n}^{\prime }}~.  \label{T2eq}
\end{align}%
To obtain these equations we used the identity $\hat{g}\hat{g}=-\hat{1}$ and
a cyclic permutation of the matrices under the trace signs.

In general, the ordinary effective vertex is to be also found by ideal
summation of the ladder diagrams incorporating residual particle-hole
interactions. Unfortunately, the Landau parameters for these interactions in
asymmetric nuclear matter are unknown therefore we simply neglect the
particle-hole interactions and consider the pair correlation function in the
BCS approximation. Thus, if the $2\times2$ matrix in spin space $\mathbf{%
\hat {\xi}}\left( \mathbf{n,}k\right) $ is some vertex of a free particle,
the ordinary vertices of a quasiparticle and a hole in the BCS approximation
are to be taken as: 
\begin{equation}
\hat{\tau}\left( \mathbf{n,}k\right) =\hat{\xi}\left( \mathbf{n,}k\right) ~,~%
\hat{\tau}^{-}\left( \mathbf{n,}k\right) =\hat{\xi}^{T}\left( -\mathbf{n,}%
k\right) ~.  \label{tau}
\end{equation}
Discarding the particle-hole interactions, we nevertheless assume that the
"bare" vertices are properly renormalized \cite{Larkin} in order to get rid
of the integration over regions far from the Fermi surface, $%
\varepsilon_{p}^{2}\gg\Delta^{2}$. As mentioned above, we omit the
renormalization factor everywhere.

\begin{figure}[h]
\includegraphics{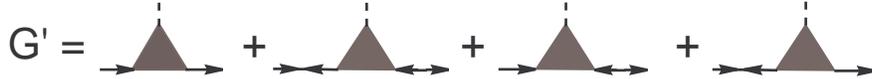}
\caption{Correction to the ordinary propagator of a quasiparticle in
external field.}
\label{fig4}
\end{figure}

Variation of the Green function of a quasiparticle under the action of
external field $U$,%
\begin{equation}
\hat{G}^{\prime }=\frac{\delta \tilde{G}}{\delta U}\mathbf{,}  \label{dG}
\end{equation}%
is given by the graphs \cite{Migdal} shown in Fig. 4, and can be written
analytically as%
\begin{align}
G^{\prime }& =GG~\hat{\tau}+FF~\left( \bm{\hat{\sigma}}\mathbf{\bar{b}}%
\right) \hat{g}\hat{\tau}^{-}\hat{g}\left( \bm{\hat{\sigma}}\mathbf{\bar{b}}%
\right)  \notag \\
& +GF~\hat{T}^{\left( 1\right) }\hat{g}\left( \bm{\hat{\sigma}}\mathbf{\bar{b%
}}\right) +FG~\left( \bm{\hat{\sigma}}\mathbf{\bar{b}}\right) \hat{g}\hat{T}%
^{\left( 2\right) }~,  \label{Gpr}
\end{align}%
where $GG\equiv G\left( p_{m}+\omega _{n},\mathbf{p+q}/2\right) G\left(
p_{m},\mathbf{p-q}/2\right) $, ect.

The medium response onto external field is given by the pair correlation
function which can be found as the analytic continuation of the following
Matsubara sum%
\begin{equation}
\Pi ^{\tau }\left( \omega _{n},q\right) =T\sum_{m}\int \frac{d^{3}\mathbf{p}%
}{8\pi ^{3}}\mathrm{Tr}\left( \hat{\tau}\hat{G}^{\prime }\right) ~.
\label{K}
\end{equation}

\section{General approach to neutrino energy losses}

The PBF processes are kinematically allowed thanks to the existence of a
superfluid energy gap, which admits the quasiparticle transitions with
time-like momentum transfer $k=\left( \omega ,\mathbf{q}\right) $, as
required by the final neutrino pair: $k=k_{1}+k_{2}$. We consider the
standard model of weak interactions. After integration over the phase space
of escaping neutrinos and antineutrinos the total energy which is emitted
into neutrino pairs per unit volume and time is given by the following
formula (see details, e.g., in Ref. \cite{L01}): 
\begin{equation}
\epsilon =-\frac{G_{F}^{2}\mathcal{N}_{\nu }}{192\pi ^{5}}\int_{0}^{\infty
}d\omega \int d^{3}q\frac{\omega \Theta \left( \omega -q\right) }{\exp
\left( \frac{\omega }{T}\right) -1}\operatorname{Im}\Pi _{\mathrm{weak}}^{\mu \nu
}\left( \omega ,\mathbf{q}\right) \left( k_{\mu }k_{\nu }-k^{2}g_{\mu \nu
}\right) ~,  \label{QQQ}
\end{equation}%
where $\mathcal{N}_{\nu }=3$ is the number of neutrino flavors; $G_{F}$ is
the Fermi coupling constant; and $\Theta \left( x\right) $ is the Heaviside
step-function. $\Pi _{\mathrm{weak}}^{\mu \nu }$ is the retarded weak
polarization tensor of the medium.

In general, the weak polarization tensor of the medium is a sum of the
vector-vector, axial-axial, and mixed terms. The mixed axial-vector
polarization has to be an antisymmetric tensor, and its contraction in Eq. (%
\ref{QQQ}) with the symmetric tensor $k_{\mu }k_{\nu }-k^{2}g_{\mu \nu }$
vanishes. Thus only the pure-vector and pure-axial polarizations should be
taken into account. We then obtain $\operatorname{Im}\Pi _{\mathrm{weak}}^{\mu \nu
}\simeq C_{\mathrm{V}}^{2}\operatorname{Im}\Pi _{\mathrm{V}}^{\mu \nu }+C_{\mathrm{%
A}}^{2}\operatorname{Im}\Pi _{\mathrm{A}}^{\mu \nu }$, where $C_{\mathrm{V}}$ and $%
C_{\mathrm{A}}$ are vector and axial-vector weak coupling constants of a
neutron, respectively.

\section{Present status of the problem}

The widely used expression for the neutrino emissivity caused by the triplet
pairing of neutrons was obtained in Ref. \cite{YKL} with the aid of the
Fermi "golden" rule. Therefore before proceeding to the self-consistent
calculation of the neutrino energy losses, it is instructive to reproduce
this formula using the calculation technique developed in our paper. We will
prove the result of Ref. \cite{YKL} can be obtained from our equations (\ref%
{QQQ}) and (\ref{K}) if to remove the field interactions through anomalous
vertices [second line in Eq. (\ref{Gpr})]. We will label the corresponding
results with tilde.

The authors of Ref. \cite{YKL} state that the weak current of
nonrelativistic neutrons is caused mostly by the temporal component of the
vector current, $\hat{J}_{0}=\Psi^{+}\hat{1}\Psi$, and by the space
components of the axial-vector current, $\hat{J}_{i}=\Psi^{+}\hat{\sigma}%
_{i}\Psi$. Consequently to reproduce their result we need to evaluate the
temporal component of the polarization tensor in the vector channel and the
spatial part of the axial polarization. Omitting the anomalous contributions
for the temporal component of the vector polarization we have to substitute
for 
\begin{equation}
\hat{\tau}=\hat{\tau}^{-}\rightarrow\hat{1}~,~\hat{T}^{\left( 1,2\right)
}\rightarrow0~,  \label{t0}
\end{equation}
where $\hat{1}$ is a unit $2\times2$ matrix in spin space. Eq. (\ref{K}) is
valid for each of the tensor components. Inserting the temporal component of
the vector vertex, from Eqs. (\ref{Gpr}), (\ref{K}) we then obtain after a
little algebra:%
\begin{equation}
\tilde{\Pi}_{\mathrm{V}}^{00}\left( \omega,q\right) =4\rho\int \frac{d%
\mathbf{n}}{4\pi}\frac{1}{2}\left( \mathcal{I}_{GG}-\bar{b}^{2}\mathcal{I}%
_{FF}\right) ~.  \label{Pi00tilde}
\end{equation}
In obtaining this expression we used Eqs. (\ref{LXX}), (\ref{IXX}) and the
identity $\left( \bm{\hat{\sigma}}\mathbf{\bar{b}}\right) \left( %
\bm{\hat{\sigma}}\mathbf{\bar{b}}\right) =\bar{b}^{2}$.

Only small transferred momenta, $q<\omega \sim T$, contribute into the
neutrino energy losses. Since the transferred momentum comes in the
polarization function in a combination $qV_{F}\ll \omega ,\Delta $ (Fermi
velocity $V_{F}$ is small in a nonrelativistic system), to the lowest
accuracy, we may evaluate the polarization tensor in the limit $\mathbf{q}=0$%
. (In the same approximation the above authors evaluate the matrix elements
of a quasiparticle transition.) Then using Eqs. (\ref{GGmFF}) and (\ref%
{IFFq0}) we find%
\begin{equation}
\operatorname{Im}\tilde{\Pi}_{\mathrm{V}}^{00}\left( \omega >0,q=0\right) =-4\pi
\rho \int \frac{d\mathbf{n}}{4\pi }\bar{b}^{2}\Delta ^{2}\frac{\Theta \left(
\omega -2\bar{b}\Delta \right) }{\omega \sqrt{\omega ^{2}-4\bar{b}^{2}\Delta
^{2}}}\tanh \frac{\omega }{4T}~.  \label{IPitilde}
\end{equation}

Polarization tensor in the axial channel can be evaluated in the same way.
In this case, omitting the anomalous contributions we have to take%
\begin{equation}
\hat{\tau}\left( \mathbf{n,}k\right) \rightarrow\hat{\sigma}_{i}~,~\hat {\tau%
}^{-}\left( \mathbf{n,}k\right) \rightarrow\hat{\sigma}_{i}^{T}~,~\hat{T}%
^{\left( 1,2\right) }\rightarrow0~.  \label{ta}
\end{equation}
Then we find after some algebraic manipulations: 
\begin{equation}
\tilde{\Pi}_{\mathrm{A}}^{ij}\left( \omega,q\right) =4\rho \int \frac{d%
\mathbf{n}}{4\pi}\left( \frac{1}{2}\left( \mathcal{I}_{GG}-\bar {b}^{2}%
\mathcal{I}_{FF}\right) \delta_{ij}~+\mathcal{I}_{FF}~\bar{b}_{i}\bar{b}%
_{j}\right)  \label{PiAtilde}
\end{equation}
In obtaining this we used the identities $\hat{g}\bm{\hat{\sigma}}^{T}\hat {g%
}=\bm{\hat{\sigma}}$, and$~\bm{\hat{\sigma}}\left( \bm{\hat{\sigma}}\mathbf{%
\bar{b}}\right) =2\mathbf{\bar{b}}-\left( \bm{\hat{\sigma}}\mathbf{\bar{b}}%
\right) \bm{\hat{\sigma}}$\textbf{.}

With the aid of Eqs. (\ref{GGmFF}) and (\ref{IFFq0}) we find:%
\begin{equation}
\operatorname{Im}\tilde{\Pi}_{\mathrm{A}}^{ij}\left( \omega >0,q=0\right) =-4\pi
\rho \int \frac{d\mathbf{n}}{4\pi }\left( \delta _{ij}-\frac{\bar{b}_{i}\bar{%
b}_{j}}{\bar{b}^{2}}\right) \bar{b}^{2}\Delta ^{2}\frac{\Theta \left( \omega
^{2}-4\bar{b}^{2}\Delta ^{2}\right) }{\omega \sqrt{\omega ^{2}-4\bar{b}%
^{2}\Delta ^{2}}}\tanh \frac{\omega }{4T}  \label{ImPiAtilde}
\end{equation}%
Inserting the imaginary part of the polarization tensor into Eq. (\ref{QQQ})
we calculate the contraction of $\operatorname{Im}\tilde{\Pi}_{\mathrm{weak}}^{\mu
\nu }$ with the symmetric tensor $k_{\mu }k_{\nu }-k^{2}g_{\mu \nu }$ to
obtain%
\begin{align}
& \operatorname{Im}\tilde{\Pi}_{\mathrm{weak}}^{\mu \nu }\left( k_{\mu }k_{\nu
}-k^{2}g_{\mu \nu }\right)   \notag \\
& =-4\pi \rho \int \frac{d\mathbf{n}}{4\pi }\bar{b}^{2}\Delta ^{2}\frac{%
\Theta \left( \omega -2\bar{b}\Delta \right) }{\mathrm{2}\omega \sqrt{\omega
^{2}-4\bar{b}^{2}\Delta ^{2}}}\tanh \frac{\omega }{4T}  \notag \\
& \times \left[ C_{\mathrm{V}}^{2}\left( q_{\parallel }^{2}+q_{\perp
}^{2}\right) +C_{\mathrm{A}}^{2}\left( 2\left( \omega ^{2}-q_{\parallel
}^{2}\right) -q_{\perp }^{2}\right) \right] ~,  \label{contr}
\end{align}%
where $q_{\parallel }$ and $q_{\perp }$ are defined as%
\begin{equation}
q_{\parallel }^{2}=\frac{1}{\bar{b}^{2}}\left( \mathbf{q\bar{b}}\right)
^{2}~,~q_{\perp }^{2}=q^{2}-q_{\parallel }^{2}~.  \label{qsq}
\end{equation}
After a little algebra we obtain the neutrino emissivity in the form:

\begin{equation}
\tilde{\epsilon}=\frac{G_{F}^{2}\mathcal{N}_{\nu }}{120\pi ^{5}}p_{F}M^{\ast
}\int \frac{d\mathbf{n}}{4\pi }\Delta _{\mathbf{n}}^{2}\int_{2\Delta _{%
\mathbf{n}}}^{\infty }d\omega \frac{\omega ^{5}}{\left( 1+\exp \frac{\omega 
}{2T}\right) ^{2}}\frac{1}{\sqrt{\omega ^{2}-4\Delta _{\mathbf{n}}^{2}}}%
\left( C_{V}^{2}+2C_{A}^{2}\right) ~,  \label{epstilde}
\end{equation}%
where $\Delta _{\mathbf{n}}\equiv \Delta \,\bar{b}\left( \mathbf{n}\right) $.

With the aid of the change $\omega =2T\sqrt{x^{2}+\Delta _{\mathbf{n}%
}^{2}/T^{2}}$ one can recast this expression to the form obtained in Ref. 
\cite{YKL}:%
\begin{equation}
\tilde{\epsilon}=\epsilon _{YKL}\equiv \frac{4G_{F}^{2}\mathcal{N}_{\nu }}{%
15\pi ^{5}}p_{F}M^{\ast }\left( C_{V}^{2}+2C_{A}^{2}\right) T^{7}\int \frac{d%
\mathbf{n}}{4\pi }\frac{\Delta _{\mathbf{n}}^{2}}{T^{2}}\int_{0}^{\infty }dx%
\frac{z^{4}}{\left( 1+\exp z\right) ^{2}}~,  \label{YKL}
\end{equation}%
where $z=\sqrt{x^{2}+\Delta _{\mathbf{n}}^{2}/T^{2}}$.

Apparently the contribution of the vector channel in this expression is a
subject of inconsistency, since conservation of the vector current in weak
interactions requires $\omega\Pi_{\mathrm{V}}^{00}\left( \omega,q\right)
=q_{i}\Pi_{\mathrm{V}}^{i0}\left( \omega,q\right) ,$ and thus one should
expect $\Pi_{\mathrm{V}}^{00}\left( \omega>0,q=0\right) =0$ for the correct
result instead of Eq. (\ref{IPitilde}). This however was not proved
explicitly for the case of triplet pairing. We now focus on this calculation.

\section{Anomalous contributions}

\subsection{Vector channel}

The self-consistent longitudinal polarization function $\Pi _{\mathrm{V}%
}^{00}\left( \omega >0,\mathbf{q}\right) $ incorporates the anomalous
contributions. At finite transferred space momentum the problem of determining the vertex corrections is much complicated. Typically massless Goldstone modes that arise due to symmetry breaking play a crucial role in conservation of the vector current. In the anisotropic $^3P_2$ phase rotational symmetry is broken and three Goldstone modes arise (termed angulons in Ref. \cite{BRS}). However, since we are interested in the specific case of $\mathbf{q}=0$
the temporal component of the anomalous vertex $\hat{T}_{\mu }$ $\left( \mu
=0,1,2,3\right) $ can be retrieved from the Ward identity which requires 
\cite{Migdal}, \cite{L08}: 
\begin{equation}
\omega \hat{T}_{0}^{\left( 1,2\right) }\left( \mathbf{n;}\omega ,\mathbf{q}%
\right) -\mathbf{q\hat{T}}^{\left( 1,2\right) }\left( \mathbf{n;}\omega ,%
\mathbf{q}\right) =\pm 2\hat{D}\left( \mathbf{n}\right) ~.  \label{Ward}
\end{equation}%
From this identity we immediately find%
\begin{equation}
\hat{T}_{0}^{\left( 1\right) }\left( \mathbf{n;}\omega ,\mathbf{q=0}\right) =%
\frac{2\Delta }{\omega }\mathbf{\bar{b}}\bm{\hat{\sigma}}\hat{g}~,
\label{T0}
\end{equation}%
and\ 
\begin{equation}
\hat{T}_{0}^{\left( 2\right) }\left( \mathbf{n;}\omega ,\mathbf{q=0}\right)
=-\frac{2\Delta }{\omega }\hat{g}\mathbf{\bar{b}}\bm{\hat{\sigma}}\mathbf{~.}
\label{T0dag}
\end{equation}

In the BCS approximation, the ordinary scalar vertices are to be taken, as
given by Eq. (\ref{t0}). Inserting the above vertices into Eqs. (\ref{Gpr}),
(\ref{K}) we obtain after a little algebra:%
\begin{equation}
\Pi _{\mathrm{V}}^{00}\left( \omega ,\mathbf{q}=0\right) =4\rho \int \frac{d%
\mathbf{n}}{4\pi }\left( \frac{1}{2}\left( \mathcal{I}_{GG}-\bar{b}^{2}%
\mathcal{I}_{FF}\right) ~+\frac{2\Delta }{\omega }\bar{b}^{2}\mathcal{I}%
_{FG}~\right) _{\mathbf{q}=0}~.  \label{Pi00}
\end{equation}%
Using Eqs. (\ref{gf}), (\ref{ff1}) yielding%
\begin{equation}
\mathcal{I}_{FG}=\frac{\omega +\mathbf{qv}}{2\Delta }\mathcal{I}_{FF}~,
\label{IFG}
\end{equation}%
we finally find%
\begin{equation}
\Pi _{\mathrm{V}}^{00}\left( \omega ,\mathbf{q}=0\right) =2C_{V}^{2}\rho
\int \frac{d\mathbf{n}}{4\pi }\left( \mathcal{I}_{GG}+\bar{b}^{2}\mathcal{I}%
_{FF}\right) _{\mathbf{q}=0}~.  \label{PI00}
\end{equation}%
Comparing this with Eq. (\ref{GGpFF}) we obtain $\Pi _{\mathrm{V}%
}^{00}\left( \omega ,\mathbf{q}=0\right) =0$, as is required by the current
conservation condition. We found that the neutrino emissivity through the
vector channel vanishes in the limit $\mathbf{q}=0$. This proves explicitly
that the neutrino emissivity via the vector channel, as obtained in Eq. (\ref%
{YKL}), is a subject of inconsistency.

\subsection{Axial channel}

We now focus on the axial channel of the weak polarization. The order
parameter in the triplet superfluid varies under the action of axial-vector
external field. Therefore the self-consistent axial polarization tensor also
must incorporate anomalous contributions. Then from Eqs. (\ref{Gpr}), (\ref%
{K}) we obtain after simple algebraic manipulations:%
\begin{align}
\Pi _{\mathrm{A}}^{ij}\left( \omega \right) & =4\rho \int \frac{d\mathbf{n}}{%
4\pi }\left[ \frac{1}{2}\left( \mathcal{I}_{GG}-\bar{b}^{2}\mathcal{I}%
_{FF}\right) \delta _{ij}+\bar{b}^{2}\mathcal{I}_{FF}\frac{\bar{b}_{i}\bar{b}%
_{j}}{\bar{b}^{2}}\right.  \notag \\
& \left. -\frac{\omega }{2\Delta }\mathcal{I}_{FF}\frac{1}{4}\mathrm{Tr}%
\left( \hat{\sigma}_{i}\hat{T}_{j}^{\left( 1\right) }\hat{g}\left( %
\bm{\hat{\sigma}}\mathbf{\bar{b}}\right) -~\hat{\sigma}_{i}\left( %
\bm{\hat{\sigma}}\mathbf{\bar{b}}\right) \hat{g}\hat{T}_{j}^{\left( 2\right)
}\right) \right] ~,  \label{KA}
\end{align}%
As in above, we focus on the case $\mathbf{q}=0$ and omit for brevity the
dependence on $\mathbf{n}$ and $\omega $. The anomalous axial-vector
vertices $\hat{T}_{j}^{\left( 1,2\right) }$ $\left( j=1,2,3\right) $ are to
be found from Eqs. (\ref{T1eq}), (\ref{T2eq}), where the ordinary vertices
are given by Eq. (\ref{ta}).

Up to this point we have not discussed the $\mathbf{n}$ dependence of $%
\mathbf{b}_{lm_{j}}\left( \mathbf{n}\right) $. This makes Eq. (\ref{PI00})
valid in the case of tensor forces resulting in the\ $^{3}P_{2}-~^{3}F_{2}$
mixing, because the general form of Eqs. (\ref{T1eq}), (\ref{T2eq}) for the
anomalous vertices takes into account not only spin-orbit interactions but
the tensor interactions in the channel of two quasiparticles.  Now we simplify 
the problem according to approximation adopted in simulations of neutron star 
cooling \cite{Page09} and consider the case of paring in the $^{3}P_{2}$ 
channel, when $l=1$, and $V_{ll^{\prime }}=\delta _{ll^{\prime }}V$, 
and the vectors $\mathbf{b}_{m_{j}}\left( \mathbf{n}\right)$ are given by
\begin{align}
\mathbf{b}_{0}& =\sqrt{\frac{1}{2}}\left( -n_{1},-n_{2},2n_{3}\right) ~, 
\notag \\
\mathbf{b}_{1}& =-\mathbf{b}_{-1}^{\ast }=-\sqrt{\frac{3}{4}}\left(
n_{3},in_{3},n_{1}+in_{2}\right) ~,  \notag \\
\mathbf{b}_{2}& =\mathbf{b}_{-2}^{\ast }=\sqrt{\frac{3}{4}}\left(
n_{1}+in_{2},in_{1}-n_{2},0\right) \ ,  \label{bm}
\end{align}%
where $n_{1}=\sin \theta \cos \varphi ,~n_{2}=\sin \theta \sin \varphi
,~n_{3}=\cos \theta $. From now on we will drop the subscript $l=1$ by
assuming $\mathbf{b}_{m_{j}}\equiv \mathbf{b}_{1,m_{j}},~\Delta _{m}\equiv
\Delta _{1,m_{j}}$, etc.

We will focus on the p-wave condensation into the state $^{3}P_{2}$ with $%
m_{j}=0$ which is conventionally considered as the preferable one in the
bulk matter of neutron stars. In this case, Eq. (\ref{bbar}) implies%
\begin{equation}
\mathbf{\bar{b}}\left( \mathbf{n}\right) =\mathbf{b}_{0}\left( \mathbf{n}%
\right) ~,~\Delta =\Delta _{0}  \label{b0}
\end{equation}%
and the gap equation (\ref{gapeq}) reads%
\begin{equation}
1=-V\int \frac{d\mathbf{n}}{4\pi }\bar{b}^{2}\left( \mathbf{n}\right)
A\left( \mathbf{n}\right) .  \label{gapeq1}
\end{equation}

From Eqs. (\ref{T1eq}) and (\ref{T2eq}) we obtain the vertex equations of
the following form $\left( i=1,2,3\right) $:%
\begin{align}
\hat{T}_{i}^{\left( 1\right) }\left( \mathbf{n}\right) & =V\sum_{m_{j}}%
\bm{\hat{\sigma}}\mathbf{b}_{m_{j}}(\mathbf{n})\hat{g}\int \frac{d\mathbf{n}%
^{\prime }}{8\pi }\left[ \mathcal{I}_{GG^{-}}\mathrm{Tr}\left( \hat{g}\left( %
\bm{\hat{\sigma}}\mathbf{b}_{m_{j}}^{\ast }\right) \hat{T}_{i}^{\left(
1\right) }\right) \right.  \notag \\
& -\mathcal{I}_{FF}\mathrm{Tr}\left( \left( \bm{\hat{\sigma}}\mathbf{b}%
_{m_{j}}^{\ast }\right) \left( \bm{\hat{\sigma}}\mathbf{\bar{b}}\right) \hat{%
g}\hat{T}_{i}^{\left( 2\right) }\left( \bm{\hat{\sigma}}\mathbf{\bar{b}}%
\right) \right)  \notag \\
& \left. -\frac{\omega }{\Delta }\mathcal{I}_{FF}2i\left( \mathbf{b}%
_{m_{j}}^{\ast }\mathbf{\times \bar{b}}\right) _{i}\right] _{\mathbf{n}%
^{\prime }}~,  \label{T1i}
\end{align}%
\begin{align}
\hat{T}_{i}^{\left( 2\right) }\left( \mathbf{n}\right) & =V\sum_{m_{j}}\hat{g%
}\bm{\hat{\sigma}}\mathbf{b}_{m_{j}}^{\ast }(\mathbf{n})\int \frac{d\mathbf{n%
}^{\prime }}{8\pi }\left[ \mathcal{I}_{G^{-}G}\mathrm{Tr}\left( \left( %
\bm{\hat{\sigma}}\mathbf{b}_{m_{j}}\right) \hat{g}\hat{T}_{i}^{\left(
2\right) }\right) \right.  \notag \\
& -\mathcal{I}_{FF}\mathrm{Tr}\left( \left( \bm{\hat{\sigma}}\mathbf{b}%
_{m_{j}}\right) \left( \bm{\hat{\sigma}}\mathbf{\bar{b}}\right) \hat{T}%
_{i}^{\left( 1\right) }\hat{g}\left( \bm{\hat{\sigma}}\mathbf{\bar{b}}%
\right) \right)  \notag \\
& \left. -\frac{\omega }{\Delta }\mathcal{I}_{FF}2i\left( \mathbf{b}_{m_{j}}%
\mathbf{\times \bar{b}}\right) _{i}\right] _{\mathbf{n}^{\prime }}~.
\label{T2i}
\end{align}%
In obtaining the last line in these equations we used $\bm{\hat{\sigma}}%
\left( \bm{\hat{\sigma}}\mathbf{\bar{b}}\right) =2\mathbf{\bar{b}}-\left( %
\bm{\hat{\sigma}}\mathbf{\bar{b}}\right) \bm{\hat{\sigma}}$ along with $%
\mathrm{Tr}\left( \left( \bm{\hat{\sigma}}\mathbf{b}_{m_{j}}^{\ast }\right)
\left( \bm{\hat{\sigma}}\mathbf{\bar{b}}\right) \bm{\hat{\sigma}}\right)
=2i\left( \mathbf{b}_{m_{j}}^{\ast }\mathbf{\times \bar{b}}\right) $, and
Eqs. (\ref{Leg}), (\ref{gf}).

Inspection of the equations reveals that the anomalous axial-vector vertices
can be found in the following form 
\begin{equation}
\mathbf{\hat{T}}^{\left( 1\right) }\left( \mathbf{n},\omega \right)
=\sum_{m_{j}}\mathbf{B}_{m_{j}}^{\left( 1\right) }\left( \omega \right)
\left( \bm{\hat{\sigma}}\mathbf{b}_{m_{j}}\right) \hat{g}~,  \label{T1A}
\end{equation}%
\begin{equation}
\mathbf{\hat{T}}^{\left( 2\right) }\left( \mathbf{n},\omega \right)
=\sum_{m_{j}}\mathbf{B}_{m_{j}}^{\left( 2\right) }\left( \omega \right) \hat{%
g}\left( \bm{\hat{\sigma}}\mathbf{b}_{m_{j}}^{\ast }\right) ~.  \label{T2A}
\end{equation}%
These general expressions can be simplified due to the fact that the
function $\mathcal{I}_{FF}\left( \mathbf{n};\omega \right) $ given by Eq. (%
\ref{FFq0}) is axial - symmetric, and the last (free) term, in Eqs. (\ref%
{T1i}) and (\ref{T2i}), can be averaged over the azimuth angle to give 
\begin{equation}
\int \frac{d\varphi }{2\pi }\left( \mathbf{b}_{0}^{\ast }\times \mathbf{\bar{%
b}}\right) =\int \frac{d\varphi }{2\pi }\left( \mathbf{b}_{2}^{\ast }\times 
\mathbf{\bar{b}}\right) =\int \frac{d\varphi }{2\pi }\left( \mathbf{b}%
_{-2}^{\ast }\times \mathbf{\bar{b}}\right) =0~,  \label{02cr}
\end{equation}%
and%
\begin{equation}
i\int \frac{d\varphi }{2\pi }\left( \mathbf{b}_{1}^{\ast }\times \mathbf{%
\bar{b}}\right) =-\mathbf{e}\frac{\sqrt{6}}{4}\bar{b}^{2}\allowbreak
~,~i\int \frac{d\varphi }{2\pi }\left( \mathbf{b}_{-1}^{\ast }\times \mathbf{%
\bar{b}}\right) =-\mathbf{e}^{\ast }\frac{\sqrt{6}}{4}\allowbreak \bar{b}%
^{2}~,  \label{1cr}
\end{equation}%
where $\mathbf{e}=\left( 1,-i,0\right) $ is a constant complex vector in
spin space. The following relations can be also verified by a
straightforward calculation:%
\begin{equation}
\int \frac{d\varphi }{2\pi }\mathbf{b}_{m_{j}}^{\ast }\mathbf{b}%
_{m_{j}^{\prime }}=\delta _{m_{j}m_{j}^{\prime }}\mathbf{b}_{m_{j}}^{\ast }%
\mathbf{b}_{m_{j}}~,  \label{bsq}
\end{equation}%
\begin{equation}
\int \frac{d\varphi }{2\pi }\left( \mathbf{\bar{b}b}_{m_{j}}^{\ast }\right)
\left( \mathbf{\bar{b}b}_{m_{j}^{\prime }}\right) =\delta
_{m_{j}m_{j}^{\prime }}\left( \mathbf{\bar{b}b}_{m_{j}}^{\ast }\right)
\left( \mathbf{\bar{b}b}_{m_{j}}\right) ~.  \label{bburb}
\end{equation}

Relations (\ref{02cr}) and (\ref{1cr}) allow to conclude that $\mathbf{B}%
_{0}^{\left( 1,2\right) }=\mathbf{B}_{\pm 2}^{\left( 1,2\right) }=0$, and%
\begin{equation*}
\mathbf{\hat{T}}^{\left( 1\right) }\left( \mathbf{n}\right) =\left( \mathbf{B%
}_{1}^{\left( 1\right) }\left( \bm{\hat{\sigma}}\mathbf{b}_{1}\right) +\mathbf{B}%
_{-1}^{\left( 1\right) }\left( \bm{\hat{\sigma}}\mathbf{b}_{-1}\right) \right) 
\hat{g}~,
\end{equation*}%
\begin{equation*}
\mathbf{\hat{T}}^{\left( 2\right) }\left( \mathbf{n}\right) =\hat{g}\left( 
\mathbf{B}_{1}^{\left( 2\right) }\left( \bm{\hat{\sigma}}\mathbf{b}_{1}^{\ast
}\right) +\mathbf{B}_{-1}^{\left( 2\right) }\left( \bm{\hat{\sigma}}\mathbf{b}
_{-1}^{\ast }\right) \right) ~.
\end{equation*}%
Inserting these expressions into Eqs. (\ref{T1i}) and (\ref{T2i}), taking
the traces and using the orthogonality relations (\ref{lmnorm}) along with
relations (\ref{bsq}), (\ref{bburb}), and 
\begin{equation}
\bar{b}^{2}\equiv \mathbf{b}_{0}^{2}~,~\mathbf{b}_{1}^{\ast }\mathbf{b}_{1}=%
\mathbf{b}_{-1}^{\ast }\mathbf{b}_{-1}~,  \label{n3}
\end{equation}%
\begin{equation}
\left( \mathbf{\bar{b}b}_{1}^{\ast }\right) \left( \mathbf{\bar{b}b}%
_{1}\right) =\left( \mathbf{\bar{b}b}_{-1}^{\ast }\right) \left( \mathbf{%
\bar{b}b}_{-1}\right) ~,  \label{n33}
\end{equation}%
we obtain the equations:%
\begin{eqnarray}
\mathbf{B}_{\pm 1}^{\left( 1\right) } &=&-V\int \frac{d\mathbf{n}}{4\pi }%
\left[ \mathcal{I}_{GG^{-}}\mathbf{B}_{\pm 1}^{\left( 1\right) }\left( 
\mathbf{b}_{1}\mathbf{b}_{1}^{\ast }\right) \right.  \notag \\
&&\left. -\mathcal{I}_{FF}\mathbf{B}_{\mp 1}^{\left( 2\right) }\left( \left( 
\mathbf{b}_{1}^{\ast }\mathbf{b}_{1}\right) \bar{b}^{2}-2\left( \mathbf{b}%
_{1}^{\ast }\mathbf{\bar{b}}\right) \left( \mathbf{\bar{b}b}_{1}\right)
\right) -\frac{\omega }{\Delta }\mathcal{I}_{FF}\mathbf{e}\frac{\sqrt{6}}{4}%
\bar{b}^{2}\right] ~,  \label{B1pm}
\end{eqnarray}%
and%
\begin{eqnarray}
\mathbf{B}_{\pm 1}^{\left( 2\right) } &=&-V\int \frac{d\mathbf{n}^{\prime }}{%
4\pi }\left[ \mathcal{I}_{G^{-}G}\mathbf{B}_{\pm 1}^{\left( 2\right) }\left( 
\mathbf{b}_{1}\mathbf{b}_{1}^{\ast }\right) \right.  \notag \\
&&\left. -\mathcal{I}_{FF}\mathbf{B}_{\mp 1}^{\left( 1\right) }\left( \left( 
\mathbf{b}_{1}\mathbf{b}_{1}^{\ast }\right) \bar{b}^{2}-2\left( \mathbf{b}%
_{1}\mathbf{\bar{b}}\right) \left( \mathbf{\bar{b}b}_{1}^{\ast }\right)
\right) +\frac{\omega }{\Delta }\mathcal{I}_{FF}\mathbf{e}^{\ast }\frac{%
\sqrt{6}}{4}\allowbreak \bar{b}^{2}\right] ~.  \label{B2pm}
\end{eqnarray}

Solution to Eqs. (\ref{B1pm}), (\ref{B2pm}) can be found in the form 
\begin{equation}
\mathbf{B}_{1}^{\left( 2\right) }=-\mathbf{B}_{-1}^{\left( 1\right) }~,~%
\mathbf{B}_{-1}^{\left( 2\right) }=-\mathbf{B}_{1}^{\left( 1\right) },
\label{Bm1}
\end{equation}%
where%
\begin{equation}
\mathbf{B}_{1}=\mathbf{e}~f\allowbreak \left( \omega \right) ~,~\mathbf{B}%
_{-1}=\mathbf{e}^{\ast }~f\allowbreak \left( \omega \right) ~,  \label{B1}
\end{equation}%
and the function $f\allowbreak \left( \omega \right) $ satisfies the equation%
\begin{equation}
~f\allowbreak =-V\int \frac{d\mathbf{n}}{4\pi }\left[ ~\allowbreak \left( 
\mathcal{I}_{GG^{-}}+\bar{b}^{2}\mathcal{I}_{FF}\right) \left( \mathbf{b}%
_{1}^{\ast }\mathbf{b}_{1}\right) f-2\mathcal{I}_{FF}~\allowbreak \left( 
\mathbf{b}_{1}^{\ast }\mathbf{\bar{b}}\right) \left( \mathbf{\bar{b}b}%
_{1}\right) f-\frac{\omega }{\Delta }\mathcal{I}_{FF}\frac{\sqrt{6}}{4}\bar{b%
}^{2}\right] ~.  \label{f}
\end{equation}%
Using Eq. (\ref{FF}) we can rewrite this as 
\begin{equation}
~f\allowbreak =-V\int \frac{d\mathbf{n}}{4\pi }\left[ ~\allowbreak \left( A+%
\frac{\omega ^{2}}{2\Delta ^{2}}\mathcal{I}_{FF}\right) \left( \mathbf{b}%
_{1}^{\ast }\mathbf{b}_{1}\right) f-2\mathcal{I}_{FF}~\allowbreak \left( 
\mathbf{b}_{1}^{\ast }\mathbf{\bar{b}}\right) \left( \mathbf{\bar{b}b}%
_{1}\right) f-\frac{\omega }{\Delta }\mathcal{I}_{FF}\frac{\sqrt{6}}{4}\bar{b%
}^{2}\right] ~.  \label{ff}
\end{equation}%
At this point it is convenient to recast the left side of this equation
according to Eq. (\ref{gapeq1}): 
\begin{equation}
f\allowbreak =-Vf\allowbreak \int \frac{d\mathbf{n}}{4\pi }\bar{b}^{2}\left( 
\mathbf{n}\right) A\left( \mathbf{n}\right) .  \label{fgap}
\end{equation}%
In this way we obtain the equation%
\begin{align}
& f\int \frac{d\mathbf{n}}{4\pi }~\allowbreak \left[ \left( \mathbf{b}%
_{1}^{\ast }\mathbf{b}_{1}-\bar{b}^{2}\right) A+2\left( \frac{\omega ^{2}}{%
4\Delta ^{2}}\left( \mathbf{b}_{1}^{\ast }\mathbf{b}_{1}\right) -~\left( 
\mathbf{b}_{1}^{\ast }\mathbf{\bar{b}}\right) \left( \mathbf{\bar{b}b}%
_{1}\right) \right) \mathcal{I}_{FF}\right]  \notag \\
& =\sqrt{\frac{3}{2}}\frac{\omega }{2\Delta }~\int \frac{d\mathbf{n}}{4\pi }%
\bar{b}^{2}\mathcal{I}_{FF}~~.  \label{feq}
\end{align}%
Since the function $\mathcal{I}_{FF}\left( \mathbf{n};\omega \right) $ is
axial - symmetric and 
\begin{equation}
\bar{b}^{2}=\frac{1}{2}\left( 1+3n_{3}^{2}\right) ~,~\mathbf{b}_{1}^{\ast }%
\mathbf{b}_{1}=\allowbreak \frac{3}{4}\left( 1+n_{3}^{2}\right) ~,
\label{bbc}
\end{equation}%
\begin{equation}
\left( \mathbf{\bar{b}b}_{1}^{\ast }\right) \left( \mathbf{\bar{b}b}%
_{1}\right) =\frac{3}{8}n_{3}^{2}\left( 1-n_{3}^{2}\right) ~,  \label{b1b}
\end{equation}%
Eq. (\ref{feq}) can be integrated over the azimuth angle, yielding the
following solution%
\begin{equation}
f~\allowbreak \left( \omega ,q=0\right) =~\frac{1}{\chi \left( \omega
,q=0\right) }~\sqrt{\frac{3}{2}}\frac{\omega }{2\Delta }\int_{0}^{1}dn_{3}%
\frac{1}{2}\left( 1+3n_{3}^{2}\right) \mathcal{I}_{FF}\left( n_{3},\omega
,T\right) ~.  \label{fEq}
\end{equation}%
where%
\begin{eqnarray}
\chi \left( \omega ,q=0\right) &\equiv &\int_{0}^{1}dn_{3}\left[ \frac{1}{4}%
\left( 1-3n_{3}^{2}\right) A\left( n_{3},T\right) \right.  \notag \\
&&\left. +\frac{3}{4}\left( \frac{\omega ^{2}}{2\Delta ^{2}}\allowbreak
\left( 1+n_{3}^{2}\right) -~n_{3}^{2}\left( 1-n_{3}^{2}\right) \right) 
\mathcal{I}_{FF}\left( n_{3},\omega ,T\right) \right] ~,  \label{hi}
\end{eqnarray}%
and the functions $A\left( n_{3},T\right) $ and $\mathcal{I}_{FF}\left(
n_{3},\omega ,T\right) $ are given by Eqs. (\ref{An}) and (\ref{FFq0}).

Explicit evaluation of Eq. (\ref{fEq}) for arbitrary values of $\omega $ and 
$T$ appears to require numerical computation. However, we can get a clear
idea of the behavior of this function using the angle-averaged energy gap in
the quasiparticle energy, $\left\langle \Delta ^{2}\bar{b}^{2}\right\rangle
\equiv \Delta ^{2}$. (Replacing angle-dependent quantities in the gap
equation with their angular average has been found to be a good
approximation \cite{Baldo}.) In this approximation the functions $\mathcal{I}%
_{FF}\left( \omega ,T\right) $ and $A\left( T\right) $, in Eqs. (\ref{fEq})
and (\ref{hi}), can be moved beyond the integrals. Using also the fact that 
\begin{equation}
A\int_{0}^{1}dn_{3}\left( 1-3n_{3}^{2}\right) =0~  \label{Am}
\end{equation}%
we find%
\begin{equation}
f=\sqrt{\frac{3}{2}}\frac{\Delta \omega }{\omega ^{2}-\Delta ^{2}/5}
\label{fappr}
\end{equation}%
Thus, in approximation of the average gap, the function $f \left( \omega
\right) $ is real-valued and is independent of the temperature.

Poles of the vertex function correspond to collective eigen-modes of the
system. Therefore, the pole at $\omega ^{2}=\Delta ^{2}/5$ signals the
existence of collective spin oscillations. The decay of the collective
oscillations into neutrino pairs gives the additive contribution into
neutrino energy losses. However, examination of the collective modes
deserves a separate study, which is beyond the scope of this paper. Here we
concentrate on the PBF processes discussed in the introduction.

In this case $\omega >2\Delta \bar{b}\left( \theta \right) \geq \sqrt{2}%
\Delta $ and, to obtain a simple analytic approximation, we omit a small
term $\Delta ^{2}/5$ in the denominator of Eq. (\ref{fappr}), thus obtaining
the axial-vector anomalous vertices in the following simple form:%
\begin{equation}
\mathbf{\hat{T}}^{\left( 1\right) }\left( \mathbf{n}\right) =\sqrt{\frac{3}{2%
}}\frac{\Delta }{\omega }\left( \mathbf{e}\left( \mathbf{\hat{\sigma}b}%
_{1}\right) +\mathbf{e}^{\ast }\left( \mathbf{\hat{\sigma}b}_{-1}\right)
\right) \hat{g}~,  \label{T1f}
\end{equation}%
\begin{equation}
\mathbf{\hat{T}}^{\left( 2\right) }\left( \mathbf{n}\right) =~\sqrt{\frac{3}{%
2}}\frac{\Delta }{\omega }\hat{g}\left( \mathbf{e}\left( \mathbf{\hat{\sigma}%
b}_{1}\right) +\mathbf{e}^{\ast }\left( \mathbf{\hat{\sigma}b}_{-1}\right)
\right) ~,  \label{T2f}
\end{equation}

Having obtained this simple result we can evaluate the axial polarization
function. Inserting (\ref{T1f}) and (\ref{T2f}) into Eq. (\ref{KA}) gives%
\begin{align}
\Pi _{\mathrm{A}}^{ij}\left( \omega \right) & =4\rho \int \frac{d\mathbf{n}}{%
4\pi }\left[ \frac{1}{2}\left( \mathcal{I}_{GG}-\bar{b}^{2}\mathcal{I}%
_{FF}\right) \delta _{ij}+\bar{b}^{2}\mathcal{I}_{FF}\frac{\bar{b}_{i}\bar{b}%
_{j}}{\bar{b}^{2}}\right.  \notag \\
& \left. +\left( \delta _{ij}-\delta _{i3}\delta _{j3}\right) \frac{3}{4}%
\bar{b}^{2}\mathcal{I}_{FF}\right] _{\mathbf{q}=0}~.  \label{Ka}
\end{align}%
The first line in Eq. (\ref{Ka}) can be evaluated with the aid of Eq. (\ref%
{GGmFF}). We find: 
\begin{equation}
\Pi _{\mathrm{A}}^{ij}=-4\rho \int \frac{d\mathbf{n}}{4\pi }\left( \delta
_{ij}-\frac{\bar{b}_{i}\bar{b}_{j}}{\bar{b}^{2}}-\frac{3}{4}\left( \delta
_{ij}-\delta _{i3}\delta _{j3}\right) \right) \bar{b}^{2}\mathcal{I}%
_{FF}\left( \omega ,q=0\right) ~.  \label{AKij}
\end{equation}%
Using Eq. (\ref{IFFq0}) we obtain the imaginary part of axial polarization:%
\begin{align}
& \operatorname{Im}\Pi _{\mathrm{A}}^{ij}\left( \omega >0,q=0\right)  \notag \\
& =-4\pi \rho \int \frac{d\mathbf{n}}{4\pi }\left( \delta _{ij}-\frac{\bar{b}%
_{i}\bar{b}_{j}}{\bar{b}^{2}}-\frac{3}{4}\left( \delta _{ij}-\delta
_{i3}\delta _{j3}\right) \right) \bar{b}^{2}\Delta ^{2}\frac{\Theta \left(
\omega ^{2}-4\bar{b}^{2}\Delta ^{2}\right) }{\omega \sqrt{\omega ^{2}-4\bar{b%
}^{2}\Delta ^{2}}}\tanh \frac{\omega }{4T}  \label{ImPiA}
\end{align}

\section{Self-consistent neutrino energy losses}

As we have obtained $\operatorname{Im}\Pi _{\mathrm{V}}^{\mu \nu }\left( \omega
>0,q=0\right) =0$, using Eqs. (\ref{IFFq0}) and (\ref{AKij}) we find%
\begin{align}
\operatorname{Im}\Pi _{\mathrm{weak}}^{\mu \nu }& =-\delta ^{\mu i}\delta ^{\nu j}C_{%
\mathrm{A}}^{2}4\pi \rho \int \frac{d\mathbf{n}}{4\pi }\left( \delta _{ij}-%
\frac{\bar{b}_{i}\bar{b}_{j}}{\bar{b}^{2}}-\frac{3}{4}\left( \delta
_{ij}-\delta _{i3}\delta _{j3}\right) \right)  \notag \\
& \times \bar{b}^{2}\Delta ^{2}\frac{\Theta \left( \omega ^{2}-4\bar{b}%
^{2}\Delta ^{2}\right) }{\omega \sqrt{\omega ^{2}-4\bar{b}^{2}\Delta ^{2}}}%
\tanh \frac{\omega }{4T}~.  \label{ImPi}
\end{align}%
Contraction of this tensor with $\left( k_{\mu }k_{\nu }-k^{2}g_{\mu \nu
}\right) $ gives: 
\begin{align}
& \operatorname{Im}\Pi _{\mathrm{weak}}^{\mu \nu }\left( k_{\mu }k_{\nu
}-k^{2}g_{\mu \nu }\right)  \notag \\
& =-\frac{1}{4}C_{\mathrm{A}}^{2}\left( 2\left( \omega ^{2}-q_{\parallel
}^{2}\right) -q_{\perp }^{2}\right) 4\pi \rho \int \frac{d\mathbf{n}}{4\pi }%
\bar{b}^{2}\Delta ^{2}\frac{\Theta \left( \omega -2\bar{b}\Delta \right) }{%
\omega \sqrt{\omega ^{2}-4\bar{b}^{2}\Delta ^{2}}}\tanh \frac{\omega }{4T}~,
\label{con}
\end{align}%
where 
\begin{equation}
q_{\parallel }^{2}=\frac{1}{\bar{b}^{2}}\left( \mathbf{q\bar{b}}\right)
^{2}~,~q_{\perp }^{2}=q^{2}-q_{\parallel }^{2}~.  \label{qq}
\end{equation}

The rest of the calculation is already performed in Sec. VII. The neutrino
energy losses can be written immediately after inspection of Eqs. (\ref%
{contr}) and (\ref{con}). From this comparison it is clear that in order to
obtain the correct neutrino energy losses, it is necessary to replace the
factor $\left( C_{\mathrm{V}}^{2}+2C_{\mathrm{A}}^{2}\right) $ with $\left(
1/2\right) C_{\mathrm{A}}^{2}$ in Eq. (\ref{YKL}). In this way we obtain%
\begin{equation}
\epsilon \simeq \frac{2}{15\pi ^{5}}G_{F}^{2}C_{\mathrm{A}}^{2}\mathcal{N}%
_{\nu }p_{F}M^{\ast }T^{7}\int \frac{d\mathbf{n}}{4\pi }\frac{\Delta _{%
\mathbf{n}}^{2}}{T^{2}}\int_{0}^{\infty }dx\frac{z^{4}}{\left( 1+\exp
z\right) ^{2}}~,  \label{eps}
\end{equation}%
where $\Delta _{\mathbf{n}}^{2}\equiv \Delta ^{2}\,\bar{b}^{2}\left( \mathbf{%
n}\right) =\frac{1}{2}\Delta ^{2}\left( 1+3\cos ^{2}\theta \right) $, and $z=%
\sqrt{x^{2}+\Delta _{\mathbf{n}}^{2}/T^{2}}$. Comparison of this expression
with Eq. (\ref{YKL}) shows that the neutrino energy losses caused by the $%
^{3}P_{2}$ pairing in neutron matter are suppressed by the factor 
\begin{equation}
\frac{1}{2}\frac{C_{\mathrm{A}}^{2}}{\left( C_{\mathrm{V}}^{2}+2C_{\mathrm{A}%
}^{2}\right) }\simeq 0.19  \label{suppr}
\end{equation}%
with respect to that predicted in Ref. \cite{YKL}.

For a practical usage we reduce Eq. (\ref{eps}) to the traditional form 
\begin{equation}
\epsilon \simeq 5.85\,\times 10^{20}~\left( \frac{M^{\ast }}{M}\right)
\left( \frac{p_{F}}{Mc}\right) T_{9}^{7}\mathcal{N}_{\nu }C_{\mathrm{A}%
}^{2}F_{t}~\frac{erg}{cm^{3}s}~,  \label{erg}
\end{equation}
where $M$ and $M^{\ast }$ are the effective and bare nucleon masses,
respectively; $c$ is speed of light, and 
\begin{equation}
F_{t}=\int \frac{d\mathbf{n}}{4\pi }\frac{\Delta _{\mathbf{n}}^{2}}{T^{2}}%
\int_{0}^{\infty }dx\frac{z^{4}}{\left( 1+\exp z\right) ^{2}}\text{.}
\label{Ft}
\end{equation}%
Notice the gap amplitude $\Delta \left( T\right) $ defined above is $\sqrt{2}
$ times larger than the gap amplitude $\Delta _{0}\left( T\right) $ used in
Ref. \cite{YKL} , where the same anisotropic gap $\Delta _{\mathbf{n}}$ is
written in the form $\Delta _{\mathbf{n}}^{2}=\Delta _{0}^{2}\left( 1+3\cos
^{2}\theta \right) $. However, the function $F_{t}$, defined in Eq. (\ref{Ft}%
), is independent of the particular choice of the gap amplitude, therefore
the analytic fit (B) suggested in Eq. (34) of Ref. \cite{YKL}, is valid and
can be used in practical computations.

\section{Summary and conclusion}

In this paper we have performed a self-consistent calculation of the
neutrino energy losses due to the pair breaking and formation processes in
the triplet-correlated neutron matter which is generally expected to exist
in the neutron star interior. Since the existing theory of anomalous weak
interactions in the fermion superfluid is well developed only for the case
of $^{1}S_{0}$ pairing we have generalized the corresponding equations for
the triplet pairing including the case when the attractive tensor coupling
is operative.

Exact solution of the vertex equations is much complicated because of
anisotropy of the triplet order parameter. Fortunately only small values of
the transferred space momenta are significant for the considered processes
in the nonrelativistic approximation. Therefore the weak vertices as well as
the polarization functions can be evaluated in the limit $\mathbf{q}=0$.

Before proceeding to the self-consistent calculation we reproduced the
neutrino energy losses as obtained in Ref. \cite{YKL}, using the calculation
technique developed in our paper. We have shown that the result of Ref. \cite%
{YKL} can be obtained in the BCS approximation from our equations (\ref{QQQ}%
) and (\ref{K}) if to remove the field interactions through anomalous
vertices.

The exact solution we found for the vector part of the weak polarization, $%
\Pi _{\mathrm{V}}^{00}\left(\omega >0,q=0\right)=0$, is consistent with the
current conservation condition. This general result, which is obtained
including the tensor couplings and the Fermi-liquid interactions, means that
the neutrino emissivity in the vector channel, as obtained in Ref. \cite{YKL}%
, is a subject of inconsistency.

The self-consistent consideration of the axial weak polarization is more
complicated. In this case, inclusion of the tensor forces and the
Fermi-liquid effects requires numerical computations even in the limit of $%
\mathbf{q}=0$. Therefore to obtain a simple analytic result we have
considered the $^{3}P_{2}$ pairing in the state with $m_{j}=0$ which is
conventionally considered as the preferable one in the minimal cooling
scenario of neutron stars. We have also neglected the residual particle-hole
interactions since the Landau parameters are unknown for the neutron matter
at high density.

Finally we used the self-consistent polarization functions for evaluation of
the neutrino energy losses due to PBF processes in the $^{3}P_{2}$ neutron
superfluid with $m_{j}=0$. The obtained self-consistent neutrino emissivity,
is given by Eq. (\ref{eps}). This expression needs to be compared to the
emissivity (\ref{YKL}) originally derived in Ref. \cite{YKL}, ignoring the
anomalous weak interactions. One can see the neutrino emissivity is strongly
suppressed due to the collective effects we have considered in this paper.
The suppression factor is $\left( 1/2\right) C_{\mathrm{A}}^{2}/\left( C_{%
\mathrm{V}}^{2}+2C_{\mathrm{A}}^{2}\right) \simeq0.19$.

Since the neutron $^{3}P_{2}$ pairing occurs in the core, which contains
more than 90\% of the neutron star volume, the found quenching of the
neutrino energy losses from the PBF processes can affect the minimal cooling
paradigm.

\end{document}